\documentclass[aps,prl,twocolumn,showpacs,amssymb]{revtex4}

\usepackage{graphicx}
\usepackage{bm}
\usepackage{amssymb,amsmath, mathrsfs}

\newcommand{\dd}{\mathrm d}
 
\newcommand{\tr}{{\mathrm{Tr}}}
 
\newcommand{\gs}{{\gtrless}}
 
\newcommand{\hyperG}{{\Psi}} 
\renewcommand{\Im}{{\mathrm{Im}}}

\newcommand{\act}{{\mathcal S}} 
 
\newcommand{\vF}{{v_F}} 
\newcommand{\gInt}{{V_0}}

\newcommand{\imp}{{\mathrm{imp}}}

\newcommand{\xb}{{\bar x}}

\newcommand{\preB}{{\mathcal C_1}}

\begin{document}

\title{Tunneling into Nonequilibrium Luttinger Liquid with Impurity}
\author{
St\'ephane Ngo Dinh,$^{1,2}$ Dmitry A. Bagrets,$^{3}$ and Alexander D.~Mirlin$^{1,2,3,4}$
}
\affiliation{
$^{1}$Institut f\"ur Theorie der Kondensierten Materie, Karlsruhe Institute of Technology, 76128 Karlsruhe, Germany\\
$^{2}$DFG Center for Functional Nanostructures, Karlsruhe Institute of Technology, 76128 Karlsruhe, Germany\\
$^{3}$Institut f\"ur Nanotechnologie, Karlsruhe Institute of Technology, 76021 Karlsruhe, Germany\\
$^{4}$Petersburg Nuclear Physics Institute, 188300 St.~Petersburg, Russia
}
\date{\today}
\begin{abstract}
We evaluate tunneling rates into/from a voltage biased quantum wire containing
weak backscattering defect. Interacting electrons in such a wire form a true {\it nonequilibrium} 
state of the Luttinger liquid~(LL). This state is created due to inelastic electron 
backscattering leading to the emission of nonequilibrium plasmons with typical
frequency $\hbar \omega \leq U$. The tunneling rates are split into two 
edges. The tunneling exponent at the Fermi edge is {\it positive} and 
equals that of the equilibrium LL, while the exponent at the side edge $E_F-U$ is 
{\it negative} if Coulomb interaction is not too strong.
\end{abstract}
\maketitle 


By virtue of advances in modern nanotechnology electron tunneling spectroscopy became
a powerful technique that enables to reveal electron correlations on
a mesoscale. Suppression or enhancement of the tunneling conductance at 
low bias is a signature of electron interaction in the system and is commonly called
a ``zero-bias anomaly'' (ZBA)~\cite{Altshuler85,Ingold92}. Measurements of ZBA in disordered 
metals~\cite{Valles89}, in high-mobility two-dimensional electron gases~\cite{Eisenstein91}, 
in the edges of quantum Hall systems~\cite{Chang96},
and recent measurements of magneto-tunneling in arrays of  
quantum wires~\cite{Jompol09} are milestones of this field.

Of particular interest on this way is the study of electron tunneling 
into quantum nanowires~\cite{Jompol09, Bockrath99,Yao99}.
Central to much of the fascinating physics of these one-dimensional (1D) electron systems is that 
Coulomb interaction has a dramatic effect leading to the emergence of 
the Luttinger liquid~(LL)~\cite{Giamarchi04}.
This strongly correlated state of matter is commonly described in terms of bosonic
elementary excitations.  Measurements of ZBA's in nanowires 
confirm predictions based on the LL model.

Notably, the behavior of a strongly correlated quantum system can change drastically 
when it is driven out of equilibrium. Remarkable examples include 
the Kondo phenomena~\cite{Franceschi02,Paaske06,Grobis08,Delattre09} 
and the Fermi-edge singularity problem~\cite{Abanin05, Snyman07}.
Recent experiments initiate the study of the nonequilibrium tunneling spectroscopy
of carbon nanotubes~\cite{Chen09} and quantum Hall edges~\cite{Altimiras}.

\begin{figure}[t]
\includegraphics[width=2.8in]{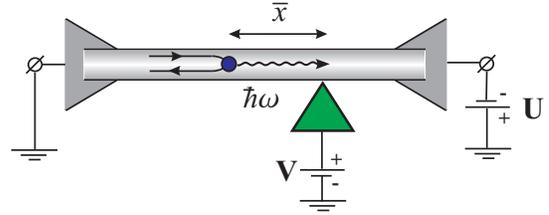}
\caption{Tunneling experiment with a voltage biased quantum wire ($\bar x>0$).
Right moving electrons have a larger chemical potential relative
to left moving electrons, $\mu_R - \mu_L = U>0$.}
\label{Device}
\end{figure}

In this paper we consider the tunneling into a voltage biased one-channel ballistic wire, 
containing a weak backscattering defect (Fig.~1).
Previous studies of this model focused on the nonlinear conductance and 
shot noise~\cite{Fendley95,Weiss96,Egger96,Trauzettel04}. However, the tunnel
spectroscopy of this problem, which requires the analysis of the single-particle Green's function,
has never been addressed.
We show that interacting electrons in such a wire form a generic nonequilibrium 
LL state, characterized by {\it non-Gaussian} plasmon correlations, and 
develop a real-time instanton approach to evaluate the tunneling rates. 
Inelastic electron backscattering at the defect induces the emission of {\it real}
nonequilibrium plasmons with typical frequencies $\hbar \omega \leq U$.
In the non-dissipative LL they transfer the shot noise of backscattered 
current with the {\it Poissonian} statistics 
to the distant point of tunneling $\bar x$  (Fig.~\ref{Device}), thereby considerably influencing the ZBA.
Let us emphasize an important difference between the present setup and that of Ref.~\cite{Gutman09}.
While in the model of~\cite{Gutman09} the nonequilibrium state is ``injected'' into the LL, here it is
created by a scatterer located {\it inside} the LL.

\begin{figure}[t]
\includegraphics[width=1.67in]{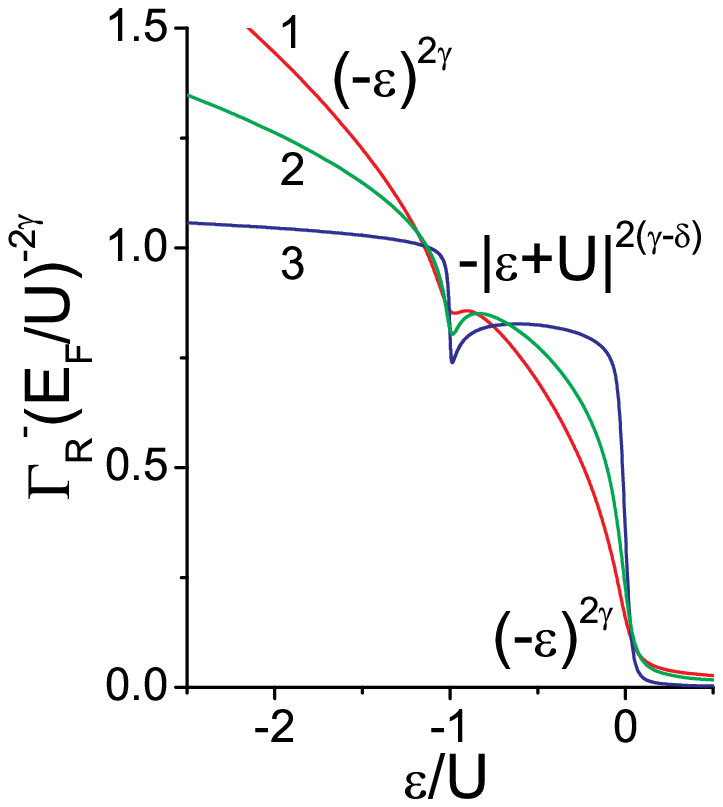}
\includegraphics[width=1.69in]{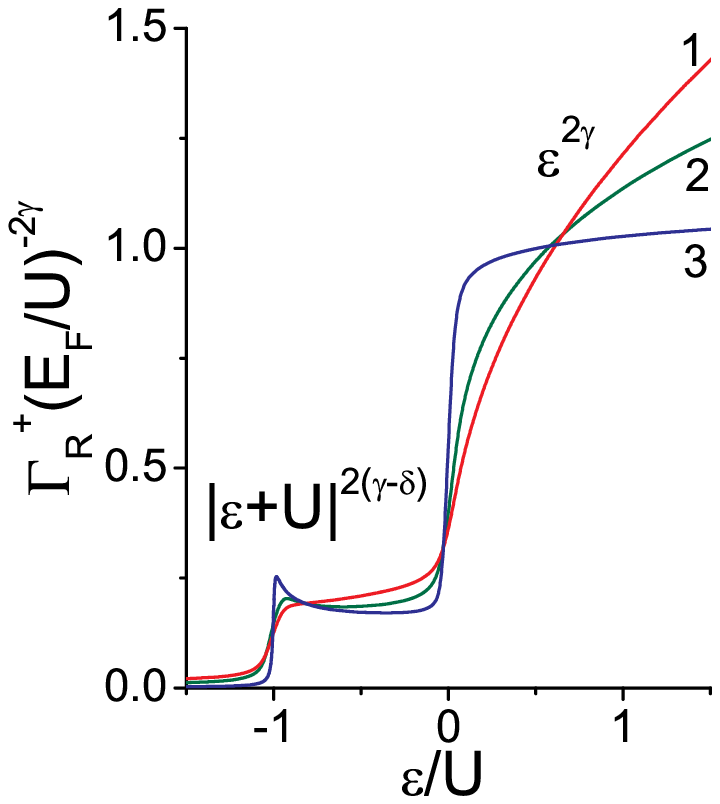}
\caption{Energy dependence of electron tunneling rates from the nonequilibrium LL (left pane) and 
into the LL (right pane), shown for $r_U^2=0.2$ and different strengths of repulsive {\it e-e} interaction: 
(1)~$K=0.4$; (2)~$K=0.5$;  and (3)~$K=0.75$. }
\label{Rates}
\end{figure}

Our results can be summarized as follows.  We consider a spinless LL with 
{\it e-e} interaction described by the conventional parameter $K$.
The tunneling rates into/from the right electron states (R-states) in the LL are split into 
two edges, $\Gamma_R^{\pm}(\epsilon) = \Gamma^{\pm}_0(\epsilon) + \Delta \Gamma^{\pm}(\epsilon)$,
as shown in Fig.~\ref{Rates}. 
The first term here accounts for the equilibrium  contribution to
ZBA around the Fermi energy,
$\Gamma_0^{\pm}(\epsilon) \propto \theta(\pm \epsilon)\bigl|\epsilon/E_F\bigr|^{2\gamma}$, with exponent $\gamma = (1-K)^2/4K$.
The nonequilibrium corrections are singular at $\epsilon = -U$: 
\begin{equation}
\Delta \Gamma_R^-(\epsilon) \propto - r_U^2\,
\bigl|\epsilon+U\bigr|^{2(\gamma-\delta)} \sin \left\{
\begin{array}{cc}
2\pi\delta, & \epsilon>-U  \\
2\pi\gamma, & \epsilon<-U  
\end{array}\right.
\label{OutR_Rate}
\end{equation}
in case of tunneling from the R-state into the tip and
\begin{equation}
\Delta \Gamma_R^+(\epsilon) \propto r_U^2\,\,\theta(\epsilon+U) 
\bigl|\epsilon+U\bigr|^{2(\gamma-\delta)}
\label{InR_Rate}
\end{equation}
in case of tunneling from the tip into the R-state.
Here $\delta=(1-K)/2$ and  $r_U^2 = r^2_0 (E_F/U)^{4\delta}/\Gamma(2K)\ll 1$ 
is the renormalized reflection coefficient~\cite{Kane92}.
The tunneling rates into/from the left electron branch remain almost equal to the equilibrium ones, 
$\Gamma_L^{\pm}(\epsilon) \simeq \Gamma^{\pm}_0(\epsilon+U)$.
The above power-law singularities are smeared on a scale
of the nonequilibrium dephasing rate
\begin{equation}
 \tau_\phi^{-1} = (2/\pi) \,U\, r_U^2 \sin^2 \pi\delta.
\label{t_phi}
\end{equation}
It is worth stressing the oscillatory dependence of $1/\tau_\phi$ on the interaction parameter $K$.
It differs from that obtained in the model of Ref.~\cite{Gutman09}, which reflects a different type of the
nonequilibrium LL state.

The result~(\ref{InR_Rate}) corresponds to 
inelastic tunneling with absorption of real plasmons. 
An electron tunneling into the LL with the energy $\epsilon<0$  can accommodate itself above 
the Fermi energy of right moving states by picking up the energy $\hbar\omega>|\epsilon|$ 
from the nonequilibrium plasmon bath. Since the energy of out-of-equilibrium plasmons
is limited by the applied voltage, one has a threshold: $\epsilon>-U$.
The correction (\ref{OutR_Rate})
describes the inverse processes --- inelastic tunneling from the LL
accompanied by the stimulated emission of nonequilibrium 
plasmons with typical energy $\hbar\omega \simeq U$. 
In the absence of interaction the splitting of tunneling rates $\Gamma^\pm_R(\epsilon)$
can be understood as the result of the double-step distribution of R-states due to the scattering off 
the impurity~\cite{Jakobs07}.

We model the wire as the spinless LL~\cite{Giamarchi04} with linear dispersion and 
short-range forward {\it e-e} 
interaction, characterized by the amplitude $V_0$, which fixes
$K=(1+\gInt/\pi\vF)^{-1/2}$. We decouple the electron field
into right- and left-moving fields $\psi_\pm$.
The wire contains a weak impurity at $x=0$ with 
a bare reflection amplitude $r_0\ll 1$.

Our theoretical analysis is based on the functional bosonization (FB)~\cite{Grishin04}.
We consider an electron motion in a Hubbard-Stratonovich
field $\varphi$ which mediates {\it e-e} interaction. 
The special feature of 1D geometry is that by a 
local gauge transformation, $\psi_\eta \mapsto e^{i \vartheta_\eta} \psi_\eta$,
the coupling between $\psi_\pm$ and $\varphi$ can be removed 
everywhere except the points of scattering
provided that $\partial_\eta \vartheta_\eta = -\varphi$, where
$\partial_\eta = \partial_t + \eta v_F\partial_x$. 
The phases $\vartheta_\pm$ define charge and current responses 
to the potential $\varphi$ in each chiral branch,  $\rho_\eta = \partial_x\vartheta_\eta/2\pi$
and $j_\eta = -\partial_t\vartheta_\eta/2\pi$. 
Following this gauge transformation one constructs the bosonized Keldysh action 
$\act = \act_b + \act_\imp$
in terms of the variables $\vartheta$ and $\phi$~\cite{bagrets08}
\begin{eqnarray}
&&\!\act_b =\!\int_{\cal C}dt\!\!\int \!\!dx \sum_{\eta = \pm} \nu\left[\frac 12
\dot\vartheta_\eta^2+\frac{\eta \vF}2
\left(\partial_x\vartheta_\eta\right)\dot\vartheta_\eta +
\varphi\dot\vartheta_\eta\right] \nonumber \\
&&\quad + \frac 1 2 \,\varphi 
\left(\gInt^{-1}+\frac1{\pi\vF}\right) \varphi, \label{Action} \\
&&\!\act_\imp  =-\frac i 4 |r_0|^2 \, \tr \Bigl[\hat g_+ e^{i \hat \Phi} 
\hat g_-e^{-i\hat \Phi}\Bigr]_{x=0}. \nonumber 
\end{eqnarray}
Here $\nu = (2\pi v_F)^{-1}$ is the 1D density of states, $\Phi=\vartheta_--\vartheta_+$,
and $\hat g_\pm$ are
``quasiclassical'' Green's function in the right/left leads. They are fixed
by boundary conditions 
$\hat g_\eta = (1-2 f_\eta)(\hat\tau_3 + i\hat\tau_2)-\hat\tau_1$,
$f_\pm(t_1-t_2)$ being the electron distribution functions. 
In the limit of zero temperature one has
$f_\pm(t) = i e^{-i\mu_\pm t}/2\pi (t + i 0)$, where
the chemical potentials satisfy 
$\mu_+-\mu_-= U$. 
The diagonal matrix
$\hat\Phi = {\rm diag}(\Phi^-, \Phi^+)$ 
and the Pauli matrices $\hat\tau$ act in the Keldysh space, 
the upper indices $\pm$ referring to the two branches of the Keldysh contour ${\cal C}$.
The trace operation (${\rm Tr}$) is performed in the Keldysh$\times$time space.
The quadratic action $\act_b$ describes the charge and current fluctuations in the clean wire,
while the impurity action $\act_\imp$ accounts for plasmon emission and 
absorption due to electron backscattering in the lowest order in $|r_0|^2$.

We start by considering Gaussian fluctuations of $\theta_\pm$ and $\varphi$,
described by the action $\act_b$. Within the FB electron phases 
have no free dynamics --- in contrast to the conventional bosonization --- 
but rather respond to the internal electric field. This response is found by optimizing
$\act_b$ for a given $\varphi$, which gives the gauge relation
$\partial_\eta \vartheta_\eta = -\varphi$. 
One has to solve it by taking the proper structure of the Keldysh 
theory into account, $\vec\vartheta_\eta[\varphi]=-\hat D_{0\eta}\hat\tau_3 \vec\varphi$.
We have introduced doublets, e.g. $\vec\varphi = (\varphi^-, \varphi^+)^T$, and the 
bare particle-hole propagator 
\begin{equation}
 D^\gs_{0\eta}(t,x) =  v_F^{-1} \,n_B^\gs(t-\eta x/\vF),
 \label{D0}
\end{equation}
where $n_B^\gs(t)= -i/2\pi(t\mp i a)$ and
$a\sim E^{-1}_F$ is a short time cut-off. 
Then the quadratic action $\act_b$, expressed
solely in terms of $\varphi$, assumes the RPA form, $\act_b=\frac12\vec\varphi^T \hat V^{-1}\vec\varphi$,
with a nonlocal effective interaction
$\hat{V} = V_0 - (V_0^2/\pi) K^2\, \partial_x\,\hat{\cal D}$.
Here $\hat{\cal D} = \frac12{\sum_{\mu\lambda}} \mu \hat{\cal D}_{\mu\lambda}$,
and $\hat{\cal D}_{\mu\lambda}$ is the propagator of the plasmon modes 
moving with velocity $u=\vF/K$,
\begin{equation}
{\cal D}^\gs_{\mu\lambda}(t,x)\!=\!\!\frac{1}{v_F}\!\!
\left[c^+_{\mu\lambda}\,n_B^\gs\!\left(t-\frac{\mu x}{u}\right)\!+\!
c^-_{\mu\lambda}\,n_B^\gs\!\left(t+\frac{\mu x}{u}\right)\right],\!
\label{Dp}
\end{equation}
where $c^+_{\mu\mu}=1+\gamma$, $c^-_{\mu\mu}=\gamma$ and $c^\pm_{\mu,-\mu}=(1-K^2)/(4K)$.

\begin{figure}[t]
\includegraphics[width=2.5in]{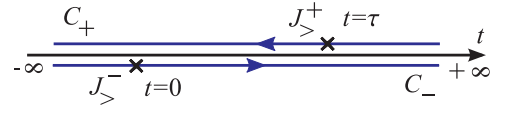}
\caption{Visualization of  $\act^{>}_J$ [ Eq.(\ref{Sj}) ] on the Keldysh contour.}
\label{Sources}
\end{figure}

To find the tunneling rates we represent 
the electron Green's function at the point of tunneling $\bar x>0$ as a path 
integral over the field $\varphi$,
\begin{equation} 
G^{\gs}_{\eta}(\bar x,\tau)\!=\!\!\int\!\mathscr{D}\varphi\, e^{i \act^{\gs}_J[\varphi] + iS[\varphi]}
G_{\eta}^{\gs}(\bar x,\tau, 0;[\varphi]).
\label{GF}
\end{equation}
Here $\hat G_{\eta}(\bar x,t,t';[\varphi])$ denotes the Green's function for a given
configuration of $\varphi$. It satisfies the Dyson equation
with the spatially local self-energy
$\hat\Sigma_\eta[\varphi]= - i\delta(x) (|r_0|^2 \vF/2) 
\,e^{i \eta \hat\Phi} \,\hat g_{-\eta}\, e^{-i \eta \hat\Phi}$
due to impurity scattering, the phase $\hat\Phi(t,[\varphi])$
being the linear functional of $\varphi$ introduced above.
The action $\act^{>}_J$ describes the creation of 
a hole at time $t=0$ and an electron at the instant $t=\tau$, 
while $\act^{<}_J$ corresponds to the inverse process,
\begin{eqnarray}
\act^{\gs}_J(\tau,\bar x,[\varphi]) = \vartheta^{\pm}_\eta(\bar x,\tau)-\vartheta^{\mp}_\eta(\bar x,0)=
 -\!\!\int\! dt\,dx \, \vec\vartheta^{\,T}_\eta \vec J_\gs.
\label{Sj}
\end{eqnarray}
The source here, e.g.
$\vec J_> = \delta(x-\bar x)\left( \delta(t), -\delta(t-\tau)\right)^T$,
acts on both branches of ${\cal C}$ as shown in Fig.~\ref{Sources}.

To find the Green's function we proceed with a semiclassical approximation~\cite{Levitov95}.
One looks for a saddle-point trajectory $\varphi^*$ which optimizes the total 
action $\act_{\rm tot} = \act + \act_J$ and further estimates the tunneling 
rate by evaluating $\act_{\rm tot}[\varphi^*]$.  
For $r_U^2\ll 1$ we can find such a trajectory 
approximately imposing that it minimizes only the quadratic part of the action, $S_0 = S_b + S_J$, 
which gives a simple linear equation in $\varphi^*$.
Taking into account corrections to $\varphi^*$ of order of $|r|^2$, 
which follow from the exact non-linear equations of motion, would lead 
to a contribution $\sim |r|^4$ to the tunneling action, 
which is beyond the accuracy of our method. 

Using the above approximation we find 
$\vec \vartheta_\mu[\varphi^*]=\hat{\mathscr D}_{\mu\eta} \vec J$.
Here the phase-phase correlation function satisfies the 
relations $\nu\,\partial_t\, \hat{\mathscr D}_{\mu\lambda} = \hat D_{0\,\mu} \delta_{\mu\lambda} - \hat{\cal D}_{\mu\lambda}$
and  $\hat{\mathscr D}_{\mu\lambda}(0,0)=0$, 
that enables easy evaluation of $\vec \vartheta^\ast$ using 
the Eqs.~(\ref{D0}) and (\ref{Dp}). 
For instance, in case of tunneling from the tip into
the right branch ($\eta=+$) the relative phases $\vec\Phi_*=\vec\vartheta^*_- - \vec\vartheta^*_+$
explicitly read
\begin{eqnarray}
&&{i \Phi_*^\pm(t)}=\ln\left[\left(\frac{t+\xb/u\mp i a}{t-\tau+ \xb/u\mp
i a}\right)^{1-\delta}\right.\label{RT_instanton}\\
&&\left.\times\left(\frac{t-\tau-  \xb/u+i a}{t- \xb/u-i a}\right)^{\delta}
\left(\frac{t-\tau+  \xb/\vF\mp i a}{t+  \xb/\vF\mp i a}\right)\right]. \nonumber
\end{eqnarray}
Substituting $\varphi^* = -\partial_\eta \vartheta_\eta^*$ into the RPA action
we obtain $i{\cal S}^\gs_0(\tau) = - 2 \gamma \ln(\pm i\tau/a)$. 
The impurity action evaluated on the instanton~(\ref{RT_instanton}) consists of four terms, 
${\cal S}^*_{\rm imp}(\tau) = \sum_{\alpha\beta} {\cal S}^{\alpha\beta}_{\rm imp}(\tau)$,
the indices $\alpha,\beta=\pm$ arising from the Keldysh structure of $\vec\Phi_*$. 
At $U>0$ the main contribution is given by
\begin{equation} 
\label{eqn:SImp}
\act_\imp^{-+} =  \frac{i|r_0|^2}{4\pi^2} \iint\!\dd^2 t\, 
\frac{e^{-i \Phi^-_*(t_1)+  i \Phi^+_*(t_2)-i U(t_1-t_2) }}
{\bigl(t_1-t_2 + i a\bigr)^{2(1-2\delta)}}.
\end{equation} 
Here we have modified the $1/t^{2}$-behavior of the bare equal point polarization operator
by taking into account the time-dependent LL renormalization
of the reflection amplitude $r(t)\sim r_0 (t/a)^{2\delta}$, which results from
quantum fluctuations around the saddle-point trajectory $\varphi^*$. 

\begin{figure}[b]
\includegraphics[width=2.6in]{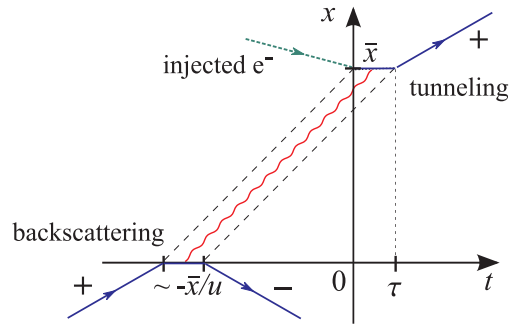}
\caption{
{\it Real} plasmons (wavy line) created in the course of inelastic electron backscattering at  
$t\sim -\bar x/u$, are absorbed by the injected electron at $t\sim 0$;
typical duration of a scattering/tunneling event is $\Delta t \sim  \tau \ll \bar x/u$. }
\label{WLines}
\end{figure}

We further concentrate on the limit of long 
$\bar x$, so that $\tau_*\ll \bar x/v_F-\bar x/u$, where
$\tau_* = {\rm max}\{\tau_\phi, 1/\Delta\epsilon\}$ is the typical 
accommodation time and $\Delta \epsilon = {\rm min}\{|\epsilon|, |U+\epsilon|\}$
is the energy relative to the nearest edge.
In this case the instanton~(\ref{RT_instanton}) consists of well separated plasmon 
and particle-hole kinks, moving with velocities $u$ and $v_F$, and
gives two independent contributions, ${\cal S}^*_{\rm p}$ and ${\cal S}^*_{\rm eh}$, to the impurity action.
Their long-time behavior at $\tau \gtrsim 1/U$ is defined
by singularities of the integrand~(\ref{eqn:SImp}):
\begin{eqnarray}
 i\act_{\rm p}^\ast(\tau) &\simeq& r_U^2\, \preB\ (i\, U\tau)^{2\delta} e^{ i\, U\tau  } - |\tau|/2\tau_\phi,
 \label{S_imp} \\
i\act_{\rm eh}^\ast(\tau) &\simeq&  \, r_U^2\, (-i\,U\tau)^{4\delta} e^{ -i \,U\tau  } \,\Gamma(2K),
\nonumber
\end{eqnarray}
where the rate  $\tau^{-1}_\phi$ is given by Eq.~(\ref{t_phi}) and the numerical factor
$\preB=\Gamma(2K)/\Gamma^2(1+\delta)$.
This asymptotics is identical in both cases of tunneling into and from the LL.
The linear growth of $\act^*_{\rm p}(\tau)$ stems from the time
domain $|t_{1,2}|\lesssim \tau$, while the oscillations $\propto \tau^{2\delta} e^{i\,U \tau}$ 
are governed by the far distant times $|t_{1,2}+\xb/u|\lesssim \tau$.
The action $\act^*_{p}$ reveals the Poissonian statistics of the shot-noise of backscattered current 
carried by plasmons (Fig.~\ref{WLines}).
Similar to that the action $\act^*_{\rm eh}$ describes the shot-noise 
due to inelastically excited electron-hole pairs at the distant times
$|t_{1,2}+\xb/v_F|\lesssim \tau$. 

Within the saddle-point approximation we obtain from the representation~(\ref{GF}) 
\begin{equation}
G_+^\gs(\bar x,\tau) \simeq -i\pi\nu \, g_+^\gs(\tau)\,e^{i S^\gs_0(\tau) + i S_p^*(\tau) + i S_{\rm ph}^*(\tau)}.
\label{GF_t}
\end{equation}
Tunneling rates are related to $G_+^\gs(\bar x,\tau)$ by the Fourier transformation.
Since $r_U^2 \ll 1$, one can expand the oscillatory part of the action in Eq.~(\ref{GF_t}),
retaining only the first term. We have checked that the particle-hole contribution $\act^*_{\rm eh}$
exactly cancels the 1st-order impurity correction to the Green's function, 
$\Delta \hat G_+ = \langle \hat G_{0,+} \hat \Sigma_+(\varphi) \hat G_{0,+} e^{i S_J} \rangle_\varphi$,
where the average is performed with the RPA-action $\act_{\rm b}$.
Keeping then only the plasmon contribution to the tunneling action, we finally find
\begin{eqnarray}
\Gamma_R^{\pm}(\epsilon) = &\pm& \frac{1}{\pi} \left({U}/{E_F}\right)^{2\gamma}\Gamma(-2\gamma)\,
\Im\left\lbrace (\mp z)^{2\gamma} \right.\label{eqn:rates}\\
	&+& \left. r_U^2\, \preB (\pm 1)^{2\gamma} 
\hyperG(-2\gamma,1+2\delta-2\gamma,-1-z)\right\rbrace, \nonumber
\end{eqnarray}
where $z=(\epsilon+i/2\tau_\phi)/U$ is the complex energy 
and $\hyperG$ is the 
confluent hypergeometric function~\cite{Gradshteyn}.
The latter is singular at $z\to -1$, yielding the power laws 
stated in Eqs.~(\ref{OutR_Rate}) and (\ref{InR_Rate}).
We plot the rates~(\ref{eqn:rates}) in Fig.~\ref{Rates} versus energy $\epsilon$ 
for different strengths of {\it e-e} interaction, indicating the edge exponents. 
Remarkably, in the vicinity of the edge $\epsilon=-U$ 
the in-rate $\Gamma_R^+$ is enhanced, while the out-rate $\Gamma_R^-$ is suppressed, 
provided the nonequilibrium exponent $\lambda=2(\gamma-\delta)$ is {\it negative}, 
which is the case of not too strong interaction realized at $K>\frac{1}{3}$.

\begin{figure}[t]
\includegraphics[width=2.6in]{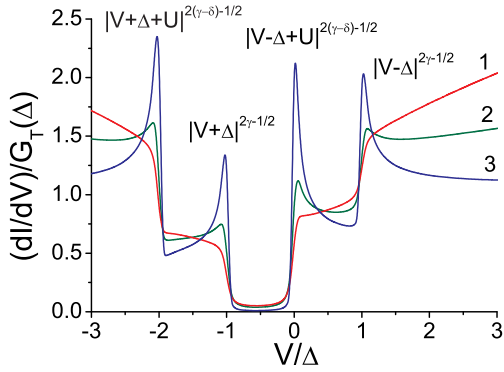}
\caption{Differential tunnel conductance 
between the BCS-type tip and the nonequilibrium LL.
The parameters are: $U=\Delta$, $r_U^2=0.2$, (1)~$K=0.4$; (2)~$K=0.5$;  and (3)~$K=0.75$.
}
\label{diffTunRates}
\end{figure}

We illustrate our theory by considering tunneling in the LL from a superconducting 
tip with the singular BCS density of states
$\nu_{\rm t}(\epsilon)\propto \lvert \epsilon\rvert(\epsilon^2-\Delta^2)^{-1/2}$.
Current measurements using this setup enable to reveal the nonequilibrium structure in 
the tunneling rates~\cite{Chen09}. For the tunneling current we have
\begin{equation}
\label{eqn:tunCurr}
I = G_T \int\!\dd \varepsilon\,\Bigl[
f_{\varepsilon-V}\,\Gamma^+_{\varepsilon}-(1-f_{\varepsilon-V})\,\Gamma^-_{\varepsilon}\Bigr]
\nu_{\rm t}(\varepsilon-V),
\end{equation}
where $\Gamma^\pm_{\varepsilon} = \Gamma_R^{\pm}(\epsilon) + \Gamma_L^{\pm}(\epsilon)$
and $G_T$ is the bare tunnel conductance. 
In Fig.~\ref{diffTunRates} we show the differential conductance $\dd I/\dd V$
in units of the normal state conductance 
$G_T(\Delta)\sim G_T (\Delta/E_F)^{2\gamma}$ at the scale  $\Delta$.
Due to the double-edge structure of the tunneling rates~(\ref{eqn:rates}) 
the peaks of the BCS density of states are split by the bias voltage $U$
and show power-law behavior with exponents $\lambda_1=2\gamma-1/2$
and $\lambda_2=2(\gamma-\delta)-1/2$. 
Singularities of $\nu_{\rm t}(\epsilon)$ visibly enhance the nonequilibrium 
structures in the rates $\Gamma^\pm_\epsilon$, making 
the conductance profile strongly asymmetric, even in the limit of small $r_U^2$.

To summarize we have developed a real-time instanton approach to the problem of
tunneling into the nonequilibrium state of the interacting quantum wire containing 
weak backscattering defect. Tunneling rates are split into two edges,
the power-law exponent $\lambda$ at the nonequilibrium edge 
$\epsilon=-U$ being {\it negative}, provided the repulsive {\it e-e} interaction is not too strong ($K>\frac{1}{3}$).
This nonequilibrium effect is associated with inelastic electron tunneling accompanied by absorption/emission 
of real plasmons with a typical frequency $\hbar\omega \sim U$.
The approach developed in this work will be useful for analysis of tunneling and interference in a
broad class of nonequilibrium LL structures with impurities and/or tunneling couplings.

We thank I.~Gornyi and D.~Polyakov for discussions. 
This work was supported by EUROHORCS/ESF and by GIF Grant No.\ 965.

\end{document}